\documentclass[a4paper,fleqn,usenatbib]{mnras}

\pdfoutput=1

\usepackage[T1]{fontenc}
\usepackage{ae,aecompl}

\usepackage{graphicx}	
\usepackage{amsmath}	
\usepackage{amssymb}
\usepackage{float}
\usepackage{array}
\usepackage[english]{babel}
\usepackage{threeparttable}
\usepackage{caption}

\defcitealias{wahlberg14}{WJJ14}
\defcitealias{wahlberg17}{WJ+17}

\newcommand{\fref}[1]{Fig. \ref{#1}}			
\newcommand{\frefp}[1]{(Fig. \ref{#1})}			
\newcommand{\frefr}[1]{(Fig. \ref{#1}, right plot)}	
\newcommand{\frefl}[1]{(Fig. \ref{#1}, left plot)}	
\newcommand{\freft}[1]{(Fig. \ref{#1}, top plot)}	
\newcommand{\frefb}[1]{(Fig. \ref{#1}, bottom plot)}	
\newcommand{\freflsee}[1]{(see Fig. \ref{#1}, left plot)}	
\newcommand{\frefrnp}[1]{Fig. \ref{#1} (right plot)}	
\newcommand{\freflnp}[1]{Fig. \ref{#1} (left plot)}	
\newcommand{\sref}[1]{Section \ref{#1}}			
\newcommand{\srefp}[1]{(Sec. \ref{#1})}			

\newcommand{\eref}[1]{Eq. \eqref{#1}}			
\newcommand{\erefp}[1]{(Eq. \ref{#1})}			
\newcommand{\erefConp}[2]{(Eqs. \ref{#1}--\ref{#2})}	
\newcommand{\erefCommp}[2]{(#2, Eq. \ref{#1})}	        
\let\d=\dsub                 
\newcommand{\d}{\textnormal{d}}
\let\t=\tsub                 
\newcommand{\t}[1]{\text{\textnormal{#1}}}  

\newcommand*\xbar[1]{%
  \hbox{%
    \vbox{%
      \hrule height 0.5pt 
      \kern0.5ex
      \hbox{%
        \kern-0.1em
        \ensuremath{#1}%
        \kern-0.1em
      }%
    }%
  }%
}

\makeatletter
\newcommand{\thickhline}{%
    \noalign {\ifnum 0=`}\fi \hrule height 1.2pt
    \futurelet \reserved@a \@xhline
}
\newcolumntype{"}{@{\hskip\tabcolsep\vrule width 1.2pt\hskip\tabcolsep}}
\makeatother

\title[Radially resolved simulations of collapsing pebble clouds]{Radially resolved simulations of collapsing pebble clouds in protoplanetary discs}

\author[K. Wahlberg Jansson and A. Johansen]{
Karl Wahlberg Jansson$^{1}$\thanks{E-mail: kalle@astro.lu.se (Lund University)}
and Anders Johansen$^{1}$
\\
$^{1}$Lund Observatory, Department of Astronomy and Theoretical Physics, Lund University, Box 43, SE-221 00 Lund, Sweden
}

\date{Accepted . Received ; in original form }

\pubyear{2017}

\begin{document}
\label{firstpage}
\pagerange{\pageref{firstpage}--\pageref{lastpage}}
\maketitle

\begin{abstract}
We study the collapse of pebble clouds with a statistical model to find the internal structure of comet-sized planetesimals. Pebble-pebble collisions occur during the collapse and the outcome of these collisions affect the resulting structure of the planetesimal. We expand our previous models by allowing the individual pebble sub-clouds to contract at different rates and by including the effect of gas drag on the contraction speed and in energy dissipation. Our results yield comets that are porous pebble-piles with particle sizes varying with depth. In the surface layers there is a mixture of primordial pebbles and pebble fragments. The interior, on the other hand, consists only of primordial pebbles with a narrower size distribution, yielding higher porosity there. Our results imply that the gas in the protoplanetary disc plays an important role in determining the radial distribution of pebble sizes and porosity inside planetesimals.
\end{abstract}


\section{Introduction}\label{sec:intro}

Planet formation occurs in protoplanetary discs surrounding newborn stars by the growth of, initially, $\mu$m-sized solid dust and ice particles up to planets with characteristic sizes $10^4-10^5$ km \citep{safronov69}. Through collisions and coagulation by contact forces small dust particles can grow to larger and larger sizes \citep[][]{blum08}. At some size ($\sim$mm-cm) the sticking capabilities of the particles become too poor for coagulation and collisions rather result in bouncing and compactification. Instead, growth can continue by concentration of pebbles in turbulent gas and the formation of gravitationally bound pebble clouds \citep[see review by][]{johansen14}. Such clouds can later experience a gravitational collapse and form solid planetesimals with sizes of order $\sim$10-100 km \citep{johansen15,simon16}.

\medskip
The ESA space probe Rosetta has, with its visit to and observations of the Jupiter family comet 67P/Churyumov--Gerasimenko (hereafter 67P), taught us a lot about the structure and interior of comets. Observations of the surface of 67P find a `goosebump' structure inside deep pits, which has been suggested to be primordial m-sized pebbles that formed the comet \citep{sierks15}. Images from the Philae lander find a pebble size scale closer to 1 cm \citep{mottola15} more in agreement with particle coagulation limits \citep{birnstiel12}. Observations of shape and gravity field result in a bulk density of $\sim$530 kg m$^{-3}$. Radar tomography measurements of the comet core \citep{patzold16,kofman15} suggest that 67P is homogeneous on length scales $<$3 m and very porous (70-75\% depending on assumed dust-to-ice ratio). These measurements overall support the picture that 67P is a porous pile of pebbles from the solar protoplanetary disc.

\medskip
The models of collapsing pebble clouds used in our previous work \citep[hereafter WJ+17]{wahlberg17} assume that pebble clouds are homogeneous, with uniform internal density, and non-rotating. The collapse itself is driven by loss of kinetic energy in inelastic collisions between pebbles. Loss of energy leads to contraction of the cloud and higher particle speeds when it tries to return to virial equilibrium \srefp{sec:model}. These models have resulting planetesimals with uniform particle size distribution throughout the interior, from the centre to the surface. A key result of these models is that the collapse time-scale is a function of both cloud mass and initial particle sizes \citep[hereafter WJJ14]{wahlberg14}. Because of their larger total surface area, small particles dissipate energy more efficiently than large particles, resulting in a faster collapse. The cloud mass also determines the magnitudes of pebble collision speeds. For low-mass pebble clouds, speeds never reach high enough values for fragmenting collisions to occur and the cloud collapses into a porous pebble-pile This would be the case for planetesimals up to $\sim$20 km in radius, covering the size range for most comet cores. 

However, in previous models we assumed, for simplicity, that protoplanetary disc gas can be ignored during the collapse. Particles of different sizes are affected differently when travelling through gas, so the gas should influence the details of the cloud collapse process. In this paper we extend the previous calculations by adding the effect of gas on the collapsing pebble cloud. We also expand the collapse model to allow the individual particle clouds to have different physical sizes. This allows us to calculate the interior structure of the resulting planetesimal, particularly how the pebble size distribution varies with distance from the planetesimal centre. Bouncing collisions during the cloud collapse cause porous pebbles to become more compact. This leads to resulting pebble sizes to not be the same as the initial, even without fragmenting collisions \citet{lorek16}. This evolution of particle density is, however, not taken into account in our simulations.

\medskip
The paper is organized as follows. The radially resolved pebble cloud model is described in \sref{sec:model}. In \sref{sec:gas} our implementation of the effect of disc gas is summarized. Our simulations of collapsing pebble clouds are presented in \sref{sec:sims} and the results are discussed in \sref{sec:conc}. In Appendix A we present an analytic description of the calculation of potential energies in a pebble cloud undergoing a gravitational collapse.

\section{Model}\label{sec:model}

We model the formation of a planetesimal through the collapse of a gravitationally bound pebble cloud. Such a cloud can form inside a protoplanetary disc through e.g. the streaming instability \citep{youdin05,johansen09,bai10}. In this cloud inelastic pebble-pebble collisions occur and dissipate kinetic energy from the cloud. This causes the cloud to contract and release kinetic energy to the pebbles due to the negative heat capacity property of self-gravitating systems. Increased pebble speeds and number density leads to more frequent collisions, higher energy dissipation rate and eventually a runaway collapse, the \textit{gravothermal catastrophe}. Here we present an expansion of our previous numerical model \citepalias{wahlberg14,wahlberg17} to also include an investigation of the effect of gas on the collapse process and the radial structure inside the resulting planetesimals.

\subsection{Model of a radially resolved gravitationally bound pebble cloud} \label{sec:1dModel}

In previous work \citepalias{wahlberg14,wahlberg17} we have assumed that all particles, regardless of size, follow each other as the cloud collapses. We keep track of the kinetic and potential energy of the entire cloud to find the radius, density, collision speeds etc. for the cloud and particle collisions. Effectively this approach spreads all the physical particles homogeneously over the entire cloud.

\begin{figure}
  \resizebox{8.2cm}{!}{\includegraphics{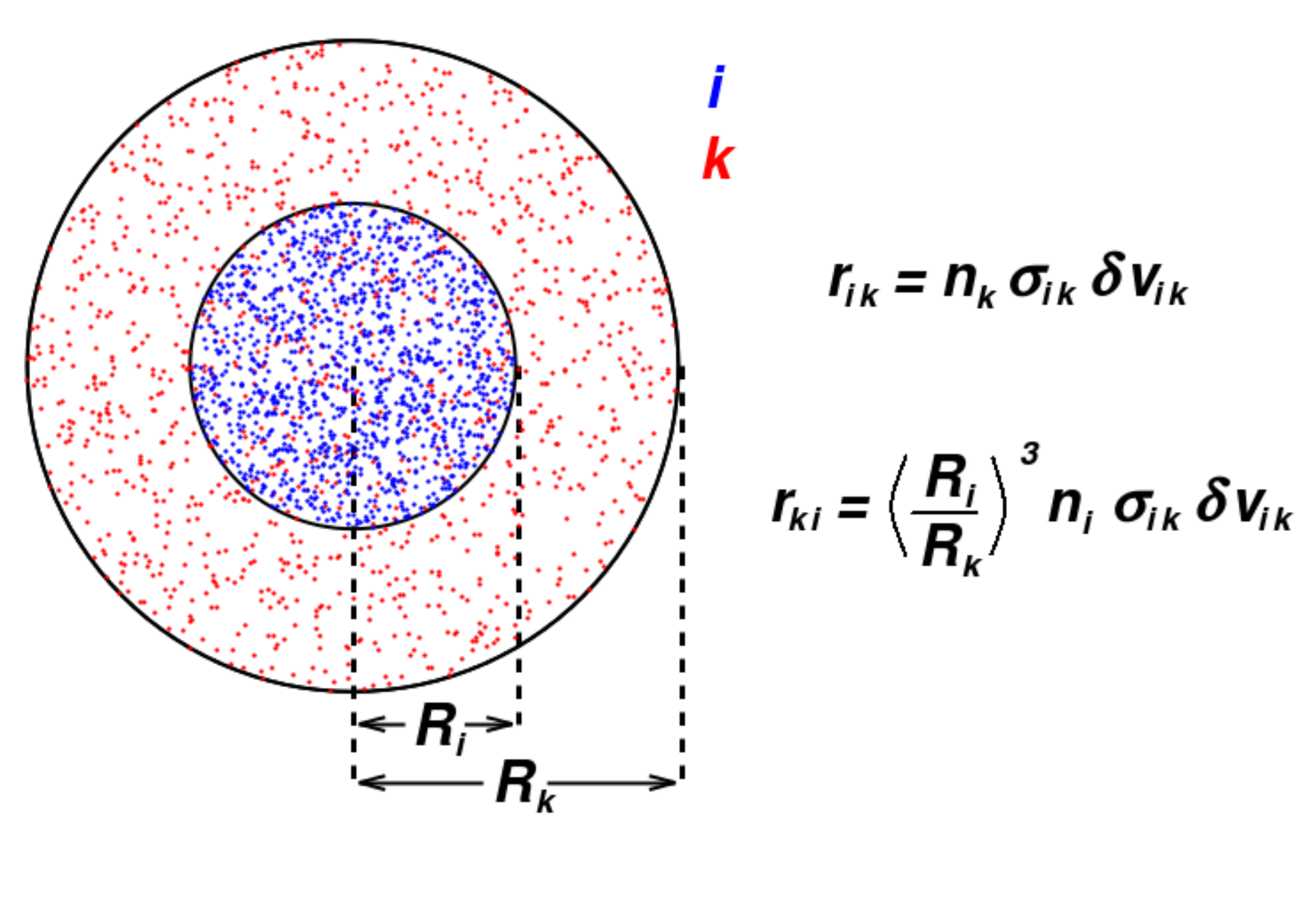}}
  \caption{Collision rates of two swarms with different cloud sizes. Only particles from swarm $k$ that are within the radius of swarm $i$, $R_i$, can collide with particles from swarm $i$.} \label{fig:subCloud}
\end{figure}

The expanded model includes the presence of the gas and the assignment of individual sizes to each collapsing sub-cloud of pebbles. The cloud is still modelled as a spherical, non-rotating cloud of pebbles but compared to the previous pebble cloud model we keep track of the radial distribution of pebbles and also allow for an initial pebble size distribution. The cloud is modelled as a number, $N_\t{p}$, of overlapping, spherical sub-clouds that together form a pebble cloud collapsing into a solid planetesimal (see \fref{fig:subCloud}). All sub-clouds are assumed to, at all times, seek virial equilibrium.

We assume that each representative particle \citep{zsom08} is its own collapsing sub-cloud and has its own kinetic energy and size to describe its properties. Using multiple sub-clouds means that representative particles are affected by the contraction of another particle swarm. The enclosed mass of one sub-cloud increases if another, larger, sub-cloud contracts (increased density). This results in a change in potential energy. We take care of this by calculating the total potential energy of each cloud and update the size of all clouds at every time step.

The collapsing cloud will eventually reach the solid density in the centre. We replace the first particle to reach solid density with a sink particle at rest in the centre of the cloud. The sink particle (core) then accretes particles from the other swarms. The core mass, $M_\t{c}$, increases continuously as

\begin{align}
 \dot{M}_\t{c}=\sum_i\pi R_\t{c}^2 v_i n_i m_i\ , \label{eq:mcDot}
\end{align}

\noindent where $R_\t{c}$ is the core radius, $v_i$ is the random particle speed in swarm $i$, $n_i$ is the particle number density and $m_i$ is the mass of those particles. This way we keep track of the particle size distribution at different depths inside the final planetesimal. Individual particle speeds are independent on the physical size of the core ($R_\t{c}$) and only at the very end of the collapse, when most of the planetesimal mass has been accreted, is the escape speed from the planetesimal larger than the virial speed of the cloud. Therefore we neglect gravitational focusing in \eref{eq:mcDot}.

\subsection{Particle collisions} \label{sec:code}

In \fref{fig:subCloud} we show the collision rates for two swarms of particles with different cloud sizes ($R_i<R_k$). The blue cloud is the denser $i$-swarm and the red cloud is the less dense $k$-swarm. The collision rate for $i$ being the representative particle is unchanged from the old model because $k$ encloses $i$ and we are only interested in the density of the $k$-swarm. The collision rate for $k$ being representative particle, on the other hand, has changed. Now we can only consider the red particles within $i$'s radius. This means that a factor $\left(R_i/R_k\right)^3$ has been added (the volume fraction of the small cloud in the large),

\begin{align}
 r_{ik} &= n_{k}\sigma_{ik}\delta v_{ik}\ , \label{eq:rik} \\ 
 r_{ki} &= \left(\frac{R_i}{R_k}\right)^3n_i\sigma_{ik}\delta v_{ik}\ . \label{eq:rki}
\end{align}

\noindent This way we can continue as before and calculate all collision rates, calculate the time until next collision, $\delta t$, and which particles collide [Eqs. (17)-(21) in \citetalias{wahlberg14}]. The outcome (coagulation, bouncing, mass transfer and fragmentation) and energy lost, $\delta E_{ik}$, in a collision as function of relative speed and particle sizes are calculated as in \citetalias{wahlberg17} with the results from experimental studies \citep{guttler10,bukhari17}. In a collision a representative particle and hence a single sub-cloud can dissipate most of its energy in a single collision (a head-on collision with low coefficient of restitution). This is unrealistic but caused no problems in the homogeneous simulations in \citetalias{wahlberg14} and \citetalias{wahlberg17}. In a collsion, a small fraction ($1/N_\t{p}$) of the particles lost kinetic energy. At the end of the time step, however, all particles were artificially given the same kinetic energy. In the new model, however, each sub-cloud keeps its energy for itself and losing all, or most, of its kinetic energy will cause free-fall of the sub-cloud after just one collision. Such a collision is unlikely but will happen when a large number of collisions occur. To resolve this `coarse-grained' behaviour we add a `smoothing parameter', $\zeta\in\left[0,1\right]$, which decreases the amount of energy dissipation per time step. The assumption, similar to the old model but implemented now for each individual sub-cloud, is that only a fraction of the particles (in the sub-cloud) collides and dissipate energy but all those particles share the same energy. This smoothing corresponds to decreasing the number of physical collisions that actually occurs during one collision between a representative and a physical particle. In these simulations, $\zeta$ is the ratio between physical particles that actually undergo a collision to the total number of physical particles belonging to the swarm of representative particle $i$. The effect of this is that both the dissipated kinetic energy and the time step decreases,

\begin{align}
 \delta E_{ik} &\rightarrow \zeta\cdot\delta E_{ik}\ , \label{eq:dESmooth} \\
 \delta t &\rightarrow \zeta\cdot\delta t\ . \label{eq:dtSmooth}
\end{align}

\noindent Another effect of smoothing is that we need to be careful when collisions result in a size change of the representative particle $i$ (coagulation, mass transfer or fragmentation). Since not all particles in swarm $i$ undergo a collision a lot of them have an unchanged size. We resolve this by making use of the Monte Carlo nature of the simulations and draw a random number uniformly distributed between 0 and 1. The probability of selecting one of the particles that has a new radius is $\zeta$. If the random number is >$\zeta$ the collision is treated as a bouncing collision with only energy dissipation as the result. Simulations of collapsing pebble clouds with different values of $\zeta$ show that the formation time of a planetesimal core varies but the overall results are unchanged \frefb{fig:sinkGrowth}. The increased resolution with smaller $\zeta$ causes the energy dissipation to be more evenly (and physically correct) spread out between sub-clouds and it takes longer for the first sub-cloud to contract into a solid core.

\subsection{Cloud energies and contraction} \label{sec:energies}

In the radially resolved model each sub-cloud corresponds to a representative particle with its own properties (e.g. cloud radius, particle size and speed, etc.). Sub-clouds do not necessarily contract with the same rate (to get into virial equilibrium) and generally do not have the same equilibrium radii, $R_{i,\t{eq}}$, as this depends on their individual kinetic and potential energy ($T_i$ and $U_i$). In virial equilibrium we have

\begin{align}
  E_{i,\t{tot}} &= -T_{i,\t{eq}} = \frac{U_{i,\t{eq}}}{2}\ , \label{eq:Vir}
\end{align}

\noindent where $E_{i,\t{tot}}$ is the current total energy of sub-cloud $i$. In the old model \citepalias{wahlberg14,wahlberg17} the potential was a function of the cloud radius only and the equilibrium radius could be analytically calculated. In the extended model the potential energy of a sub-cloud is more complex; it is the sum of the contributions from all other sub-clouds, $U_{ik}$, and the core \erefConp{eq:appU}{eq:appUik}. The expression for $U_{ik}$ is different depending on which sub-cloud, $i$ or $k$, is larger. Now we cannot find $R_{i,\t{eq}}$ analytically unless it is smaller than the innermost cloud \erefCommp{eq:Req_in}{$R_{i,\t{eq}}<R_k$ $\forall k$} or larger than the outermost cloud \erefCommp{eq:Req_out}{$R_{i,\t{eq}}>R_k$ $\forall k$}. Otherwise we do not know which sub-clouds $k$ are larger and which are smaller than $i$ at $R_{i,\t{eq}}$. This issue is solved by keeping track of the radii and potential energies of all sub-clouds in a sorted array. To, continuously, find the equilibrium radii for sub-clouds we use their total energy and \eref{eq:Vir} to get $U_{i,\t{eq}}$. Next we compare the equilibrium potential to the sorted radius and potential arrays and find at what cloud size, $R_{i,\t{eq}}$, the potential is $U_{i,\t{eq}}$. This is done for all sub-clouds every time step as a change in radius of one sub-cloud affects the potentials of all other sub-clouds.

\begin{figure*}
 \begin{center}
  \resizebox{17cm}{!}{\includegraphics{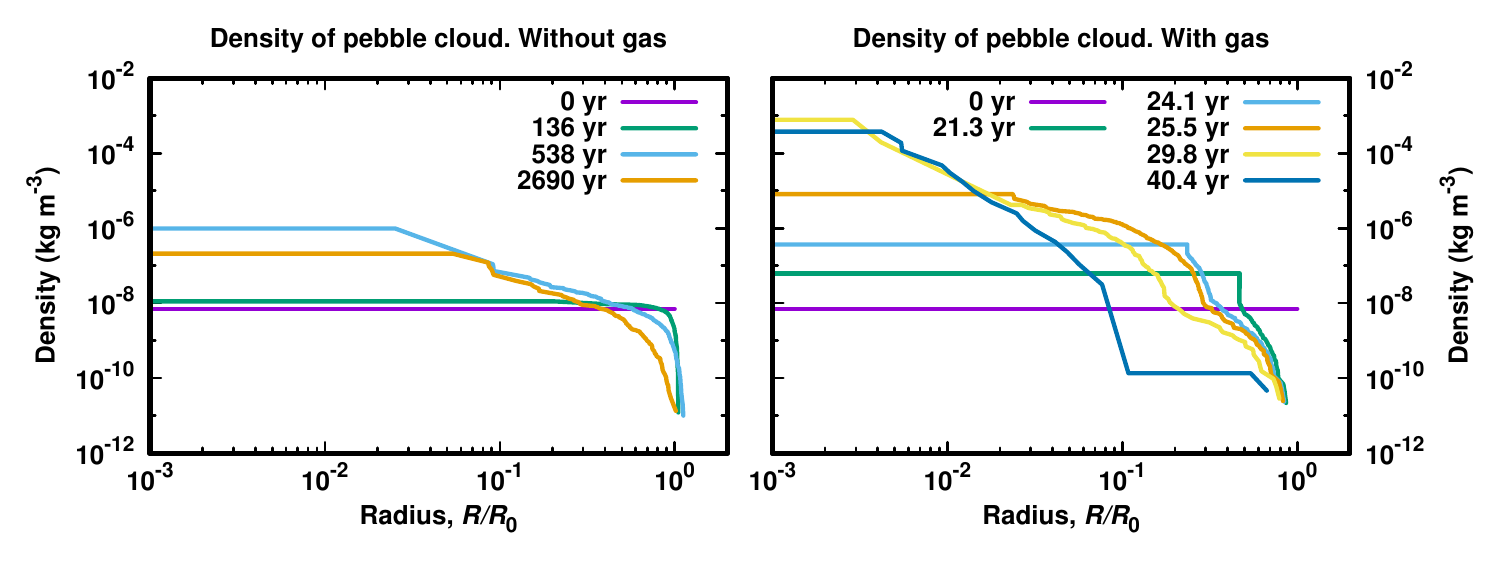}}
  \caption{Evolution of density at different depths in the cloud during collapse to a planetesimal of $R_\t{solid}=5$ km. Only the particles left in the cloud contribute to the densities shown in the plots. When a solid core forms the number of remaining pebbles in the cloud decreases, resulting in lower collision rates. Without gas (left plot) energy is only dissipated through inelastic pebble-pebble collisions. The collapse time is prolonged by the decreasing collision rates in the gradually depleted cloud (see also Fig. \ref{fig:sinkGrowth}). The inclusion of gas (right plot) results in significant energy dissipation, even when collision rates are small, and the cloud extension decreases continuously. This leads to higher densities and a faster collapse. The densities at time zero in both plots are horizontal lines because of the assumption that the pebble cloud initially is homogeneous.} \label{fig:rhoEvo}
 \end{center}
\end{figure*}

\subsection{Collapse limits} \label{sec:gas}

Self-gravitating pebble clouds collapse on a short time-scale. The collapse time decreases with increasing planetesimal size but, for cm-sized pebbles, it takes only a couple of orbital periods to form planetesimals of a few km in size \citepalias{wahlberg14}. At some cloud mass ($\sim$40 km solid planetesimal radius for cm-sized pebbles) the energy dissipation from inelastic collisions starts to become more rapid than the release of kinetic energy (virialization, \sref{sec:model}). This leads to a collapse at the free-fall speed. To resolve this in the simulations we limit the contraction speed to the free-fall speed

\begin{align}
 v_{i,\t{ff}}=\sqrt{\frac{2GM_{i,\t{enc}}}{R_{i,0}}}\sqrt{\frac{R_{i,0}-R_i}{R_i}} \label{eq:v_iff}\ ,
\end{align}

\noindent where $M_{i,\t{enc}}$ is the enclosed mass and $R_{i,0}$ is the radius of latest virial equilibrium of sub-cloud $i$ respectively. The free-fall limit results in, for massive clouds, particles achieving sub-virial velocities and the collapse becomes `cold'. One result of this is that even for the most massive pebble clouds collision speeds, in a large part of the collapse, are small enough that collisions do not result in fragmentation and some pebbles survive the entire collapse \citepalias{wahlberg14}.

\subsection{Effect of gas friction}

In the homogeneous cloud model the effect of gas on the collapse was not taken into account. We discussed this approximation in \citetalias{wahlberg17} and came to the conclusion that in early parts of the collapse gas drag should matter. In the extended model we include gas drag and limit particle speeds to their terminal speeds, $v_\t{t}$. The terminal speed for a particle in sub-cloud $i$ can be written as \citep[e.g.][]{whipple72,weidenschilling77,birnstiel10}

\begin{align}
 v_{i,\t{t}} = \tau_{i,\t{f}}F_{i,\t{g}}=\frac{\pi\rho_\t{s} a_i}{2\Sigma_\t{d}\Omega_\t{K}}\frac{GM_{i,\t{enc}}}{R_i^2}\ , \label{eq:vTerm}
\end{align}

\noindent where $\tau_{i,\t{f}}$ is the friction time of the particle, $F_{i,\t{g}}$ is the external force on the particle (gravity), $\rho_\t{s}$ is the solid density of the particle, $\Sigma_\t{d}$ and $\Omega_\t{K}$ are the surface density and Keplerian frequency of the protoplanetary disc at the cloud's distance to the star. We get \eref{eq:vTerm} for the terminal speed by assuming that the particles are small (cm-sizes) and experience Epstein drag from the gas. The change in the random speed of a particle in sub-cloud $i$ due to gas drag, $\dot{v}_{i,\t{d}}$, is calculated as

\begin{align}
 \dot{v}_{i,\t{d}} = -\frac{v_i}{\tau_{i,\t{f}}}\ . \label{eq:vDotDrag}
\end{align}

\noindent We use the minimum mass solar nebula \citep{hayashi81} to model the protoplanetary disc containing the collapsing pebble cloud and get $\Sigma_\t{d}$ to limit both the random speed of particles and the contraction speed of the cloud. Including gas into the simulations can both speed up (particles lose energy faster due to gas drag) and slow down (if the contraction speed is faster than $v_{i,\t{t}}$) a collapse.

\subsection{Algorithm summary}

The algorithm can be summarized in a series of steps done every time step:  

\begin{enumerate}
 \item Find the time step $\delta t$ and the particles included in the collision (representative particle $i$ and physical particle $k$) from the collision rate matrix $r_{ik}$ \erefConp{eq:rik}{eq:rki}.
 \item Calculate the outcome of the $ik$-collision \srefp{sec:code}. Swarm $i$ loses energy $\delta E_{ik}$. The energy lost and time step is decreased from the effect of smoothing \erefConp{eq:dESmooth}{eq:dtSmooth}.
 \item Contract all sub-clouds toward virial equilibrium using $\delta t$, potential energy and kinetic energy from the previous time step. To find the equilibrium radius of each sub-cloud we use the sorted potential array \srefp{sec:energies}. Particle speeds increase due to virialization and decrease due to gas drag \erefp{eq:vDotDrag}. These contractions are limited by free-fall and terminal speed \erefConp{eq:v_iff}{eq:vTerm}.
 \item Increase the mass of the planetesimal core (if any) and decrease the mass of all sub-clouds correspondingly \erefp{eq:mcDot}.
 \item Calculate the new enclosed masses $M_{i,\t{enc}}$ \erefConp{eq:mEnc}{eq:mEncik}, collision rates $r_{ik}$ and potential energies $U_i$ \erefp{eq:appU}.
 \item Keep track of free-fall, $v_{i,\t{ff}}$, and terminal, $v_{i,\t{t}}$, speeds for each sub-cloud so that particles do not move too fast or that the sub-clouds do not collapse too fast. Reset $v_{i,\t{ff}}$ if a sub-cloud reaches virial equilibrium.
\end{enumerate}

\section{Results}\label{sec:sims}

With our simulations we investigate the implementation of the radially resolved collapse model described in \sref{sec:model} and how the results differ from the homogeneous cloud model in \citetalias{wahlberg14} and \citetalias{wahlberg17}. We are mainly interested in how the particle size distribution varies with depth inside planetesimals. The shape of the size distribution will affect the packing capabilities and hence the overall density and porosity of the resulting planetesimal. Pebble clouds are formed in gaseous discs around newborn stars and should be embedded in gas. Therefore we also explore the effect of gas on the collapse.

\subsection{Initial conditions and simulation setup}

\begin{figure}
  \resizebox{8.2cm}{!}{\includegraphics{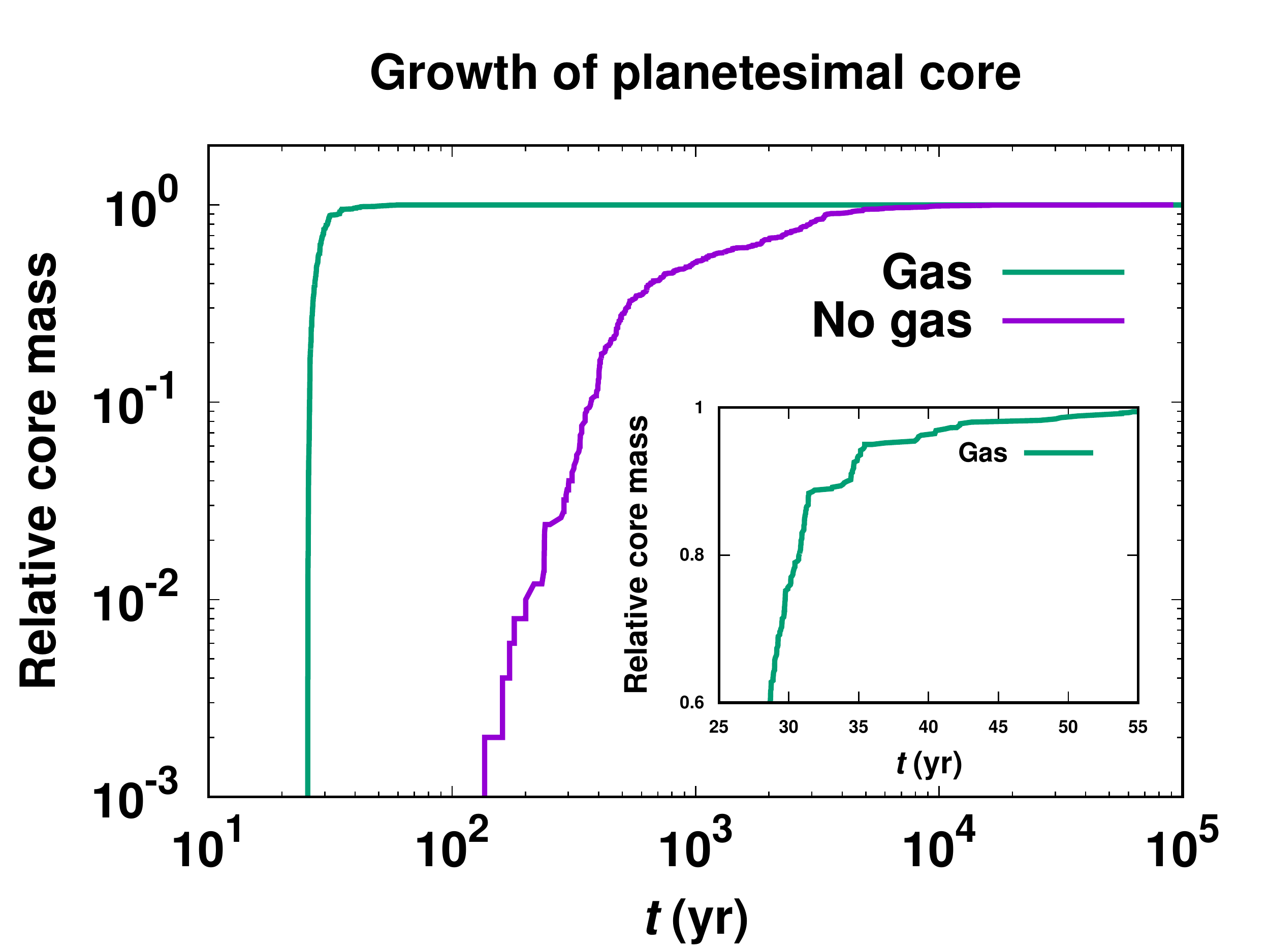}}
  \resizebox{8.2cm}{!}{\includegraphics{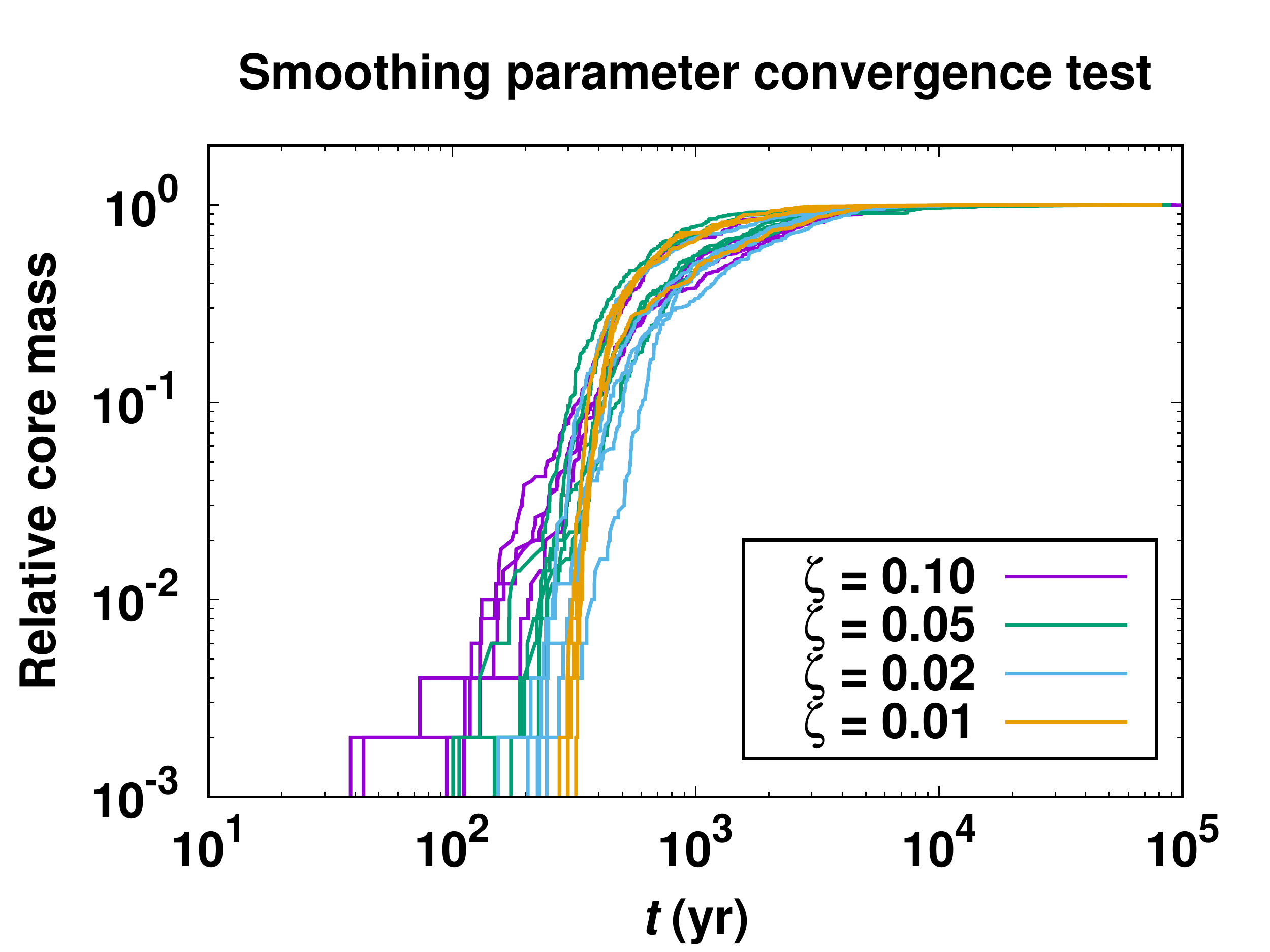}}
  \caption{Growth of planetesimal core for a planetesimal of $R_\t{solid}=5$ km (top plot). Gas causes continuous and efficient energy dissipation for the pebbles, which results in both a faster formation of a solid planetesimal core and a shorter time-scale for the entire collapse. The low cloud density at the final stages of the collapse without gas increases the collapse time-scale significantly. The bottom plot shows the core growth for simulations with different values of the `smoothing parameter' $\zeta$ \erefConp{eq:dESmooth}{eq:dtSmooth}. The formation time of the first solid core increases slightly with decreasing $\zeta$ but not the overall collapse time-scale.} \label{fig:sinkGrowth}
\end{figure}

For small clouds \citepalias[$R_\t{solid}\lesssim$30 km,][]{wahlberg17} collisions result in bouncing only so pebble sizes remain unchanged. Since the collapse time is a function of pebble size \citepalias[e.g.][]{wahlberg14} we are interested in an initial particle size distribution. Investigations of meteorites \citep[e.g. ][]{eisenhour96} suggest that the sizes of chondrules are distributed according to a log-normal or Weibull distribution. We use this as inspiration of the size distribution of pebbles concentrated by the streaming instability and forming self-gravitating clouds. In our simulations we then start with a log-normal particle size distribution (skewed slightly more toward smaller particles),

\begin{align}
\d N(a) &= \frac{1}{a\sigma_a\sqrt{2\pi}}e^{-\frac{(\ln a-\mu_a)^2}{2\sigma_a^2}} \d a\ , \label{eq:logNorm}
\end{align}

\noindent where $\mu_a$ and $\sigma_a$ is the mean and standard deviation of the logarithm of the particle radius $a$. Since we are investigating pebble clouds we set $e^{\mu_a}=1$ cm and $\sigma_a=0.5$ to study the collapse of a cloud of broadly mm-dm-sized particles.

Initially, the cloud consists a number of homogeneous, spherical, non-rotating, equal-sized sub-clouds of pebbles. It is put at 39 astronomical units from the Sun (the semi-major axis of Pluto, to investigate comet formation) with sizes of each sub-cloud set to the Hill radius of the total mass of all clouds. This choice causes both the density and free-fall time of a pebble cloud to be independent of mass. The radius of the pebbles in each individual sub-cloud is picked randomly from the log-normal size distribution described above.

\subsection{Collapsing pebble clouds} \label{sec:collapse}

The simulations investigate the formation and interiors of comet-sized (5 km) planetesimals from the collapse of gravitationally bound pebble clouds. 

\begin{figure*}
 \begin{center}
  \resizebox{17cm}{!}{\includegraphics{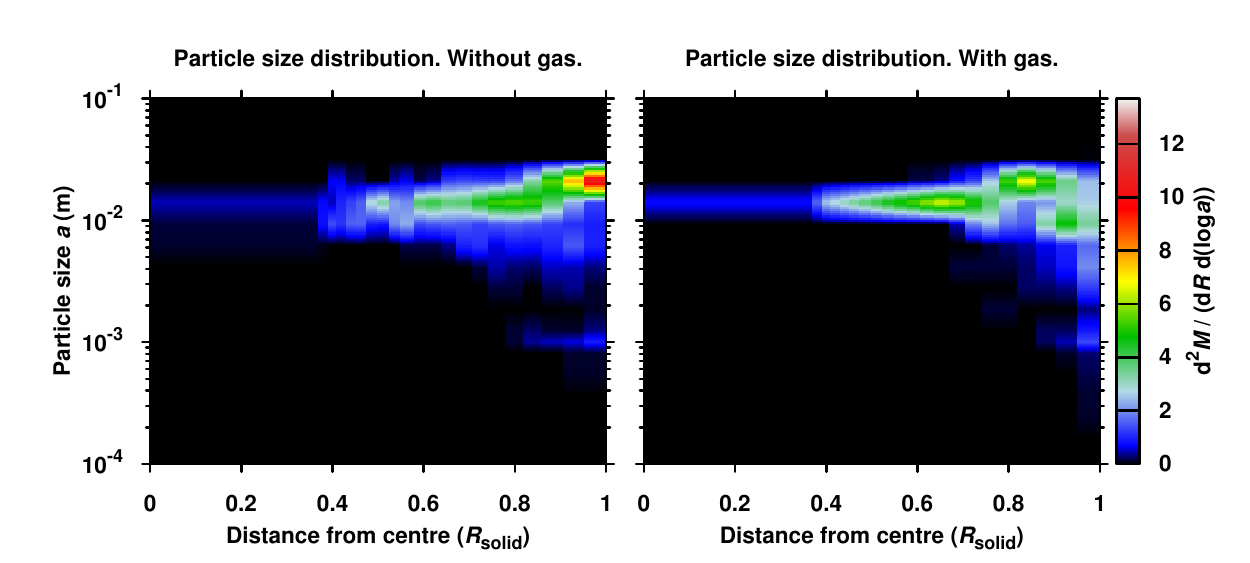}}
  \resizebox{17cm}{!}{\includegraphics{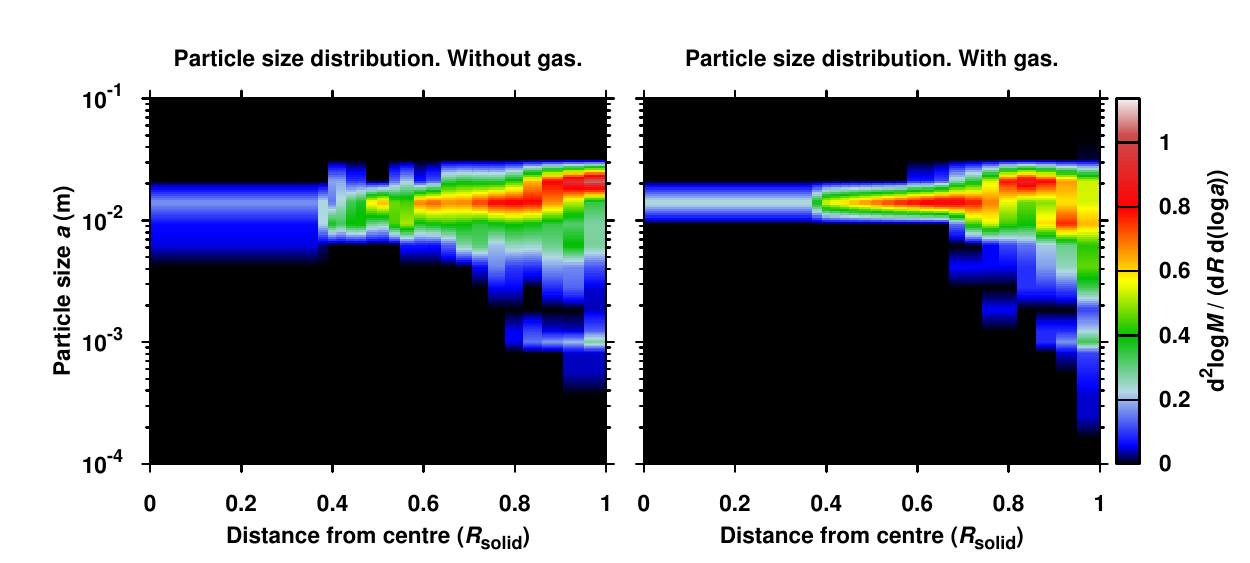}}
  \caption{Mass weighted particle size distributions at different depths inside a planetesimal of $R_\t{solid}=5$ km. Both simulations start with a log-normal particle size distribution \erefp{eq:logNorm}. Ignoring gas (left plots) results in a planetesimal with an `onion'-like interior \freflsee{fig:pebblePile} with the smallest particles in the centre since they have a more efficient energy dissipation rate \citepalias{wahlberg14}. The inclusion of gas (right plots) adds gas drag \erefp{eq:vTerm}, but also limits the collapse speed to the terminal speed of particles. Small particles fall slowly through the gas while large particles have low energy dissipation, so the first particles to collapse to solid density represent the middle of the size range (1-2 cm). These are followed by the largest particles (2-3 cm in size) and then the smallest particles ($\lesssim$1 cm) at a relative distance of 0.9 from the center \frefr{fig:pebblePile}. In the end of the collapse most of the mass is in the solid planetesimal core and the cloud density is low. This causes slower energy dissipation and higher particle speeds. The resulting fragmenting collisions can be seen by the presence of small particles (pebble fragments) in the outermost layer of the planetesimal. NB The bottom two plots are equivalent to the top two but with their colour scheme determined by the logarithm of the mass in each size bin.} \label{fig:sizeDistr}
 \end{center}
\end{figure*}

Because of higher collision rates \erefConp{eq:rik}{eq:rki} and stronger coupling to the gas \erefp{eq:vTerm} smaller particles dissipate their kinetic energy more efficiently than larger pebbles. This causes sub-clouds with different particle sizes to contract at different rates and the density inside the cloud to decrease with distance to the centre \frefp{fig:rhoEvo}. During the collapse the density increases in the inner parts of the cloud and, at some point, one sub-cloud is the first to reach solid density and will form a planetesimal core. As time goes on, this core accretes particles from the other sub-clouds and grows in mass \freft{fig:sinkGrowth}. The increase in core mass decreases the number of remaining pebbles and the density of the outer parts of the cloud which, in turn, results in lower collision rates. In simulations including gas, the particles are subject to gas drag and lose kinetic energy very efficiently even when the number of pebbles is low at the end of the collapse. This keeps densities high \frefr{fig:rhoEvo} and the collapse time short \freft{fig:sinkGrowth}. The solid centre of a planetesimal of $R_\t{solid}=5$ km is formed already after $\sim$25 yr (similar to the free-fall time) and the rest of the material has all been accreted after $\sim$50 yr. For simulations without gas, on the other hand, the particles only lose energy in dissipative collisions resulting in a slower collapse.

\begin{figure*}
 \begin{center}
  \resizebox{8.8cm}{!}{\includegraphics[trim=40mm 0mm 40mm 0mm]{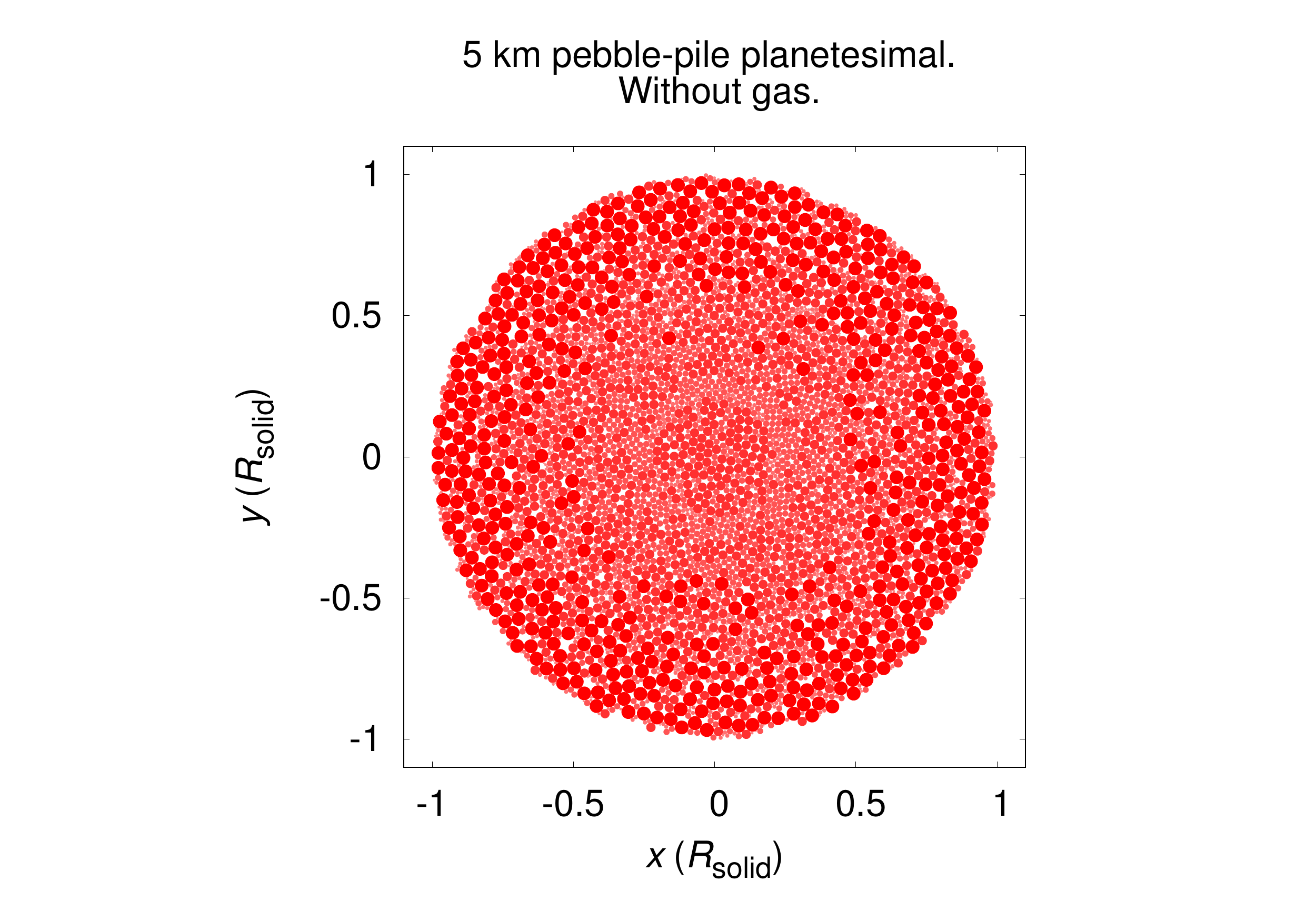}}
  \resizebox{8.8cm}{!}{\includegraphics[trim=40mm 0mm 40mm 0mm]{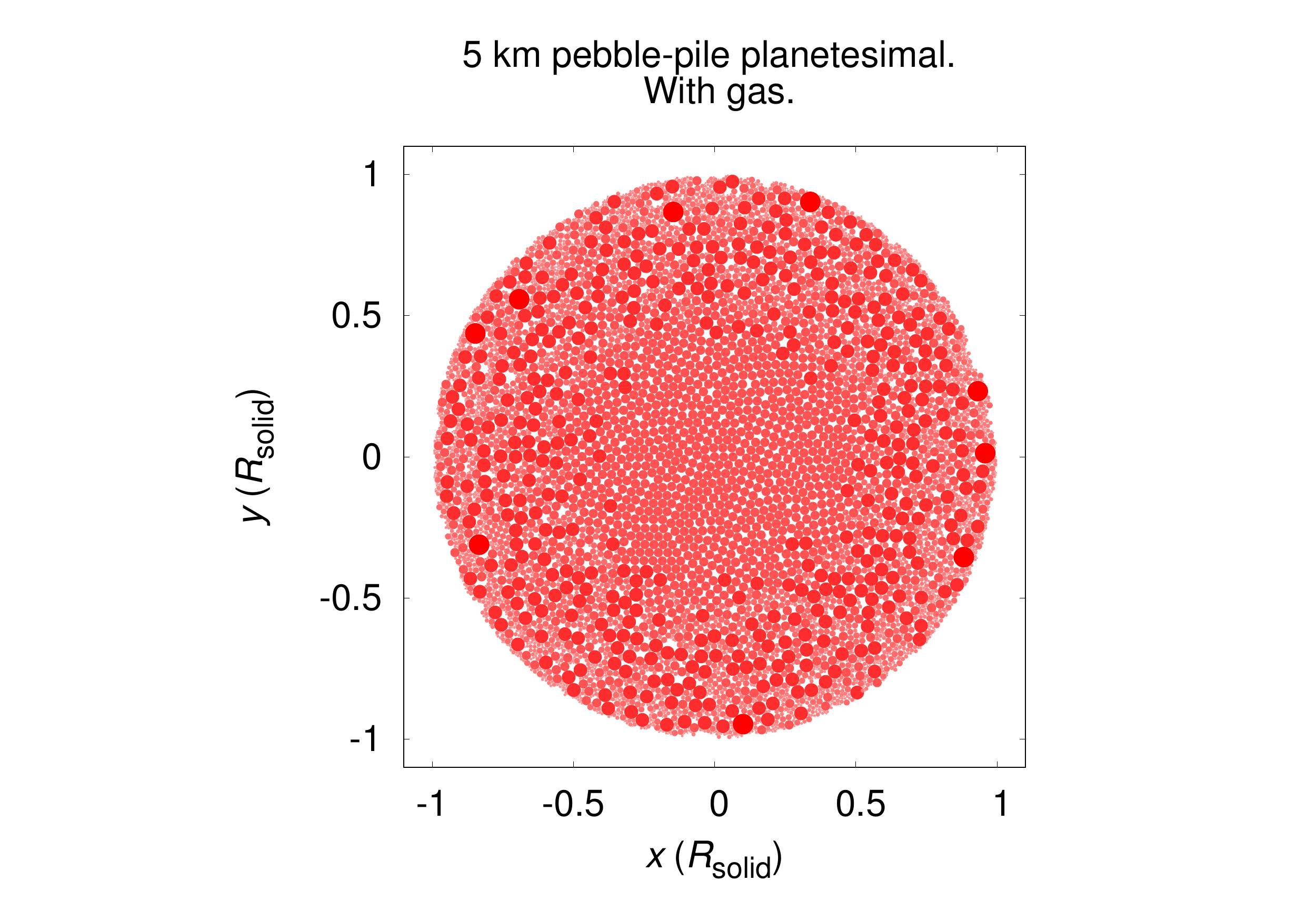}}
  \caption{Layered pebble-pile planetesimals with particle size distributions from simulations ($R_\t{solid}=5$ km, \fref{fig:sizeDistr}). The left figure shows results from a simulation without gas and the right figure results from one with the effect of gas included. Small particles are gathered in the centre of the left planetesimal since they dissipate kinetic energy in the collapse phase faster both because of the higher collision frequency. The inclusion of gas adds both gas drag and a terminal speed limit to the collapse \erefp{eq:vTerm}. This results in a planetesimal with the middle-sized particles in the centre, followed by larger particles and finally a mixture of small and large particles where fragmenting collisions have contributed to the formation of additional small particles. The algorithm for producing these plots is described in \sref{sec:collapse}. Symbol areas in figure proportional to the area of corresponding pebble. Note that pebble sizes and planetesimal size are not to scale.} \label{fig:pebblePile}
 \end{center}
\end{figure*}

\begin{figure*}
 \begin{center}
  \resizebox{17cm}{!}{\includegraphics{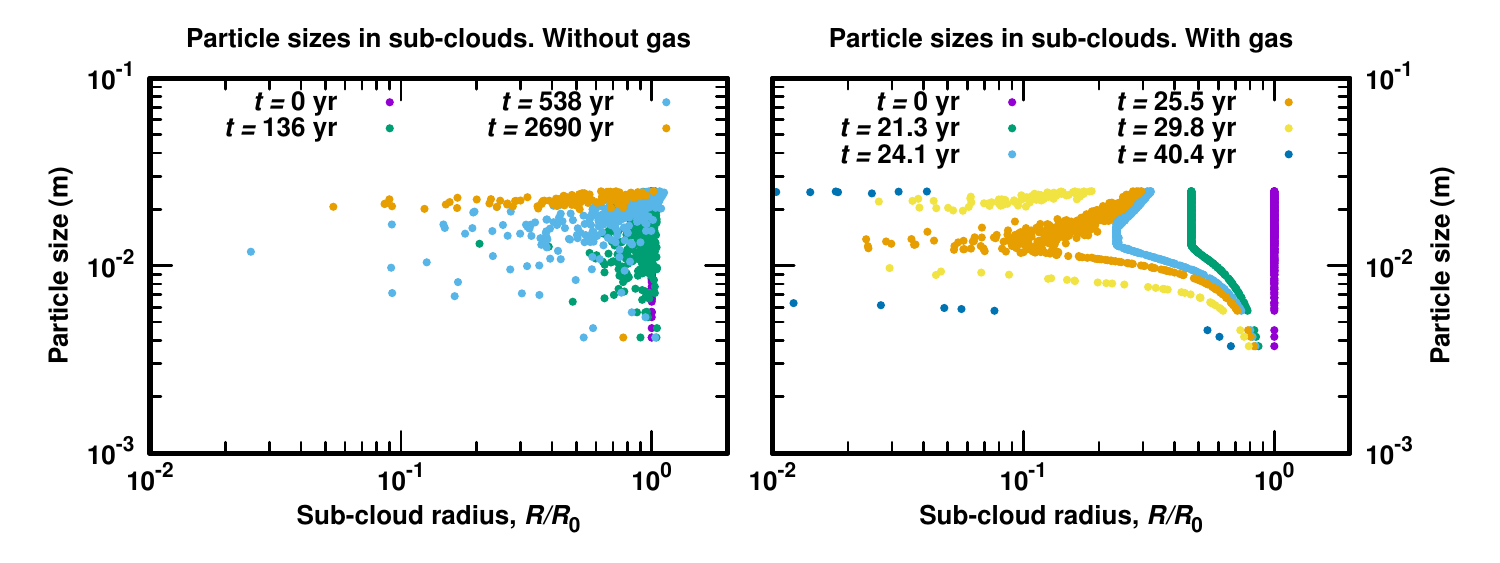}}
  \caption{Cloud size and particle size for all sub-clouds at different times in the collapse to a planetesimal of $R_\t{solid}=5$ km. Without gas (left plot) energy dissipation is most efficient for small particles, leading to a continuous increase in the average particle size as the planetesimal core forms and accretes mass. Including the effect of gas (right plot) causes small particles to rapidly lose energy due to efficient gas drag. However, the collapse is limited by the terminal speed of the particles which is lower for small particles \erefp{eq:vTerm}. The result of this is that mid-sized particles ($\sim$1 cm), small enough for efficient gas drag but also large enough to overcome the terminal speed limit, collapse fastest and form the planetesimal core.} \label{fig:sizeEta}
 \end{center}
\end{figure*}

An important aim of the collapse simulations is to investigate the porosity inside a pebble-pile planetesimal from the particle packing capabilities. This is done by keeping track of particle size distributions at different depths as material is accreted onto the growing planetesimal core. \fref{fig:sizeDistr} shows the particle size distributions at different depths inside resulting planetesimals with radii $R_\t{solid}=5$ km, one plot for a simulation neglecting gas (left plot) and one with the effect of gas included (right plot). By using these size distributions and a two-dimensional circular version of the drop-and-roll algorithm of \citet{hitti13} we produce pebble-pile plots (\fref{fig:pebblePile}) to investigate the packing capabilities inside the planetesimals. These plots show the size distribution of pebbles at proper distances from the centre but the porosity in two dimensions does not mirror the porosity in three dimensions. In three dimensions the packing mechanism results in a different porosity. The plots are produced by continuously generating angles randomly between 0 and $2\pi$, drop pebbles toward the centre of the planetesimal from those angles and let them roll to lowest potential (gravity pointing towards the centre). The sizes of the pebbles depends on the current size of the planetesimal and are generated from the size distributions in \fref{fig:sizeDistr}. \fref{fig:pebblePile} shows that including gas into the simulations has both positive and negative effects on the collapse for pebbles of different sizes. In the case of simulations ignoring gas sub-clouds with small particles contract faster and are the first to form the solid core of the planetesimal. As the collapse goes on, larger and larger pebbles lose their energy to land on the core and the planetesimal grows to have a somewhat layered interior \frefl{fig:pebblePile}. This can also be seen in \freflnp{fig:sizeEta}, which shows the particle size vs. cloud size for the sub-clouds at different points in the collapse. The overall trend is that the sub-clouds with large particles take longer time to contract and eventually reach solid density.

In the case of simulations including gas, however, the resulting planetesimal has a different internal structure \frefr{fig:pebblePile}. As can be seen in \frefrnp{fig:sizeEta}, the sub-clouds with mid-sized ($\sim$1 cm) pebbles collapse the fastest and form the core. Both sub-clouds with smaller and larger particles, on the other hand, take longer to reach solid density and, mixed together, form the outer layers. This is caused by the fact that gas not only dissipate energy from the pebbles but also limits the collapse speed to the terminal speed \erefp{eq:vTerm} of the particles. The sub-clouds with the smallest particles lose their kinetic energy quickly by gas drag but cannot contract faster than the terminal speed. The larger particles have a higher terminal speed ($v_\t{i,t}\propto a_\t{i}$) and the contraction is less limited by the gas. However, the gas drag on the larger particles is smaller and they lose energy less efficiently which eventually result in a slower collapse. The optimal particle size is somewhere in between ($\sim$1 cm) where the terminal speed is high enough to not hinder the collapse too much but still small enough that the gas drag is efficient.

One difference between the radially resolved (this paper) and the old homogeneous model \citepalias{wahlberg14,wahlberg17} is that fragmenting pebble-pebble collisions also occur in the formation of low-mass planetesimals. The surface layer of a resulting planetesimal contains small particles (pebble fragments) which can be seen in the particle size distributions of the outermost layer in \fref{fig:sizeDistr}. As discussed in \sref{sec:gas} the collapse becomes `cold' (sub-virial particle speeds) if the energy dissipation is too rapid. This helps pebbles to survive collisions in the end of the collapse where the virial speed needs to be high to support the increasing gravity \erefp{eq:Vir}. The last phase of a cloud collapse consists of a solid planetesimal core surrounded by a small fraction of the total mass as a cloud with low density. The low density results in slower energy dissipation and the possibility to get closer to virial equilibrium after each collision. An increase in terminal speed \erefp{eq:vTerm} as cloud contracts allows for higher particle speeds even when the effect of gas is included. In turn, the collision speeds reach values high enough for fragmentation and small pebble fragments will land on the surface of the planetesimal. This does not happen in the old homogeneous model \citepalias{wahlberg14,wahlberg17} because there the sub-clouds collapse together resulting in efficient energy dissipation in the end of the collapse (a core with surrounding low-density cloud never forms).

\subsection{Pebble-pile comets}\label{sec:comets}

The formation of comets (and other km-sized planetesimals) through the collapse of self-gravitating pebble clouds results, as in \citepalias{wahlberg14,wahlberg17}, in porous pebble-piles. In low-mass pebble clouds collision speeds only reach values high enough for silicate dust aggregates to fragment in the very end of the collapse \frefp{fig:sizeDistr}. Ice aggregate pebbles would not fragment at any point in the collapse to small planetesimals \citepalias{wahlberg17}. Our simulations show that the environment of the initial pebble cloud is important to the internal structure of the resulting planetesimal. If the disc gas surrounding the pebbles is taken into account, the collapse process is different. Gas increases energy dissipation through drag but also limits the contraction speed of pebbles \srefp{sec:gas}. This causes a comet formed in the outer regions of the Solar System to be a pebble-pile with mid-sized ($\sim$1 cm) pebbles in the centre of the comet and surface layers made of a mixture of larger and smaller pebbles \frefr{fig:pebblePile}. A mixture of particle sizes would lead to more efficient packing and lower porosity. Without disc gas the comet has an `onion'-like interior with shells of pebbles decreasing in size with increasing depth \frefl{fig:pebblePile}. The shell structure observed for each individual lobe of 67P/Churyumov--Gerasimenko \citep{massironi15} suggests the lobes could have formed from the contraction of a pebble cloud and that activity erodes into layers of different pebble size distributions (and hence porosities). Measurements of a permittivity gradient down to $\sim$100 m inside 67P \citep{ciarletti15} implies a variation comet bulk properties. The permittivity decreases with increasing depth suggesting higher porosity and/or decrease in dust-to-ice ratio further inside the comet.

\section{Conclusions}\label{sec:conc}

In this paper we have expanded the \citet{wahlberg17} model of planetesimal formation through the collapse of gravitationally bound pebble clouds inside protoplanetary discs. The collapse results from energy dissipation in inelastic pebble-pebble collisions inside the clouds. Our previous models have assumed that particles were spread homogeneously inside the cloud. Our new model takes into account the presence of the gaseous protoplanetary disc and the radial structure of the resulting planetesimal. To investigate the effects of disc gas we run two sets of simulations, one with and one without the effects caused by gas \srefp{sec:gas}.

As suggested in the analysis of \citet{wahlberg14} the collapse process is a strong function of particle sizes. The total surface area is larger for small particles than for an equal mass in larger particles and this results in higher collision rates and more efficient energy dissipation. Small particles are also more affected by gas \erefp{eq:vTerm}, increasing the energy dissipation further. This causes the particle size distribution inside a planetesimal to be a function of depth. Most of the largest pebbles are found in the outer layers of the planetesimal \frefp{fig:sizeDistr}. In simulations ignoring gas this results in a planetesimal with an `onion'-like structure. Including gas, however, leads to a planetesimal interior with mid-sized pebbles ($\sim$1 cm) in the centre and smaller and larger particles in the outer layers. This is caused by the fact that gas not only provides drag (stronger for small particles) but also limits the contraction speed to the terminal speed of the particles \erefp{eq:vTerm}. The particle size distribution inside the planetesimal is also affected by the evolution of the density of individual pebbles.

An important effect of resolving the radial structure is that a core forms early in the collapse phase and grows by accreting the surrounding cloud. Without gas, energy is dissipated only by particle collisions and the collapse time-scale is long \freft{fig:sinkGrowth}. For the gas model, gas drag leads to continuous energy dissipation even at low densities towards the end of the collapse. At the end of the collapse phase most of the mass is in the core and the remaining cloud has a low density. The low density leads to low collision rates and the possibility to get back into virial equilibrium, leading to higher collision speeds which, in turn, cause fragmenting collisions. This results in a broader particle size distribution in the outer layers of the planetesimal \fref{fig:sizeDistr}. In previous models the collapse is cold (sub-virial velocities because of free-fall limit) so, for small planetesimals, bouncing is the only outcome of collisions.

Our results suggest that comets formed through the collapse of pebble cloud are pebble-piles with a porous interior. The sizes of pebbles will vary with depth inside the comet \frefp{fig:sizeDistr}. The resulting comet structure differs if the cloud collapsed embedded in gas or not \fref{fig:pebblePile}. Our simulations assume pebbles are silicate dust aggregates which result in possibility for fragmenting collisions at the end of the collapse. If the pebbles are made of ice, however, the collision speeds would never reach values high enough for fragmentation \citep{wahlberg17} and the range of particle sizes in the surface layer of the comet would be narrow.

Our radially resolved collapse model still lacks physical properties, such as rotation. A non-zero initial angular momentum of a pebble cloud undergoing a collapse could result in binary objects or discs forming around new planetesimals. Through N-body simulations \citet{nesvorny10} found formation of binary planetesimals in gravitational instability of pebble concentrations. This planetesimal formation mechanism matches observations of trans-Neptunian objects. Both a high fraction of binary objects and that components in a binary have the same composition have been observed \citep{noll08,benecchi09}. The next step in our investigations is to increase the dimensionality further and investigate the effect of rotation. discs should form around planetesimals which we plan to investigate with hydrodynamical simulations and present in a subsequent paper.

\section*{Acknowledgements}
KWJ and AJ were supported by the European Research Council under ERC Starting Grant agreement 278675-PEBBLE2PLANET. AJ was also supported by the Swedish Research Council (grant 2010-3710) and by the Knut and Alice Wallenberg Foundation (grants 2012.0150, 2014.0017 and 2014.0048). KWJ and AJ acknowledges the support from the Royal Physiographic Society in Lund for grants to purchase computer hardware to run the simulations on. We also thank the referee Kees Dullemond for insightful comments that helped improve the manuscript.

\bibliographystyle{mnras} 
\bibliography{myRef} 

\begin{thebibliography}{}
\makeatletter
\relax
\def\mn@urlcharsother{\let\do\@makeother \do\$\do\&\do\#\do\^\do\_\do\%\do\~}
\def\mn@doi{\begingroup\mn@urlcharsother \@ifnextchar [ {\mn@doi@}
  {\mn@doi@[]}}
\def\mn@doi@[#1]#2{\def\@tempa{#1}\ifx\@tempa\@empty \href
  {http://dx.doi.org/#2} {doi:#2}\else \href {http://dx.doi.org/#2} {#1}\fi
  \endgroup}
\def\mn@eprint#1#2{\mn@eprint@#1:#2::\@nil}
\def\mn@eprint@arXiv#1{\href {http://arxiv.org/abs/#1} {{\tt arXiv:#1}}}
\def\mn@eprint@dblp#1{\href {http://dblp.uni-trier.de/rec/bibtex/#1.xml}
  {dblp:#1}}
\def\mn@eprint@#1:#2:#3:#4\@nil{\def\@tempa {#1}\def\@tempb {#2}\def\@tempc
  {#3}\ifx \@tempc \@empty \let \@tempc \@tempb \let \@tempb \@tempa \fi \ifx
  \@tempb \@empty \def\@tempb {arXiv}\fi \@ifundefined
  {mn@eprint@\@tempb}{\@tempb:\@tempc}{\expandafter \expandafter \csname
  mn@eprint@\@tempb\endcsname \expandafter{\@tempc}}}

\bibitem[\protect\citeauthoryear{{Bai} \& {Stone}}{{Bai} \&
  {Stone}}{2010}]{bai10}
{Bai} X.-N.,  {Stone} J.~M.,  2010, \mn@doi [\apj]
  {10.1088/0004-637X/722/2/1437}, \href
  {http://adsabs.harvard.edu/abs/2010ApJ...722.1437B} {722, 1437}

\bibitem[\protect\citeauthoryear{{Benecchi}, {Noll}, {Grundy}, {Buie},
  {Stephens}  \& {Levison}}{{Benecchi} et~al.}{2009}]{benecchi09}
{Benecchi} S.~D.,  {Noll} K.~S.,  {Grundy} W.~M.,  {Buie} M.~W.,  {Stephens}
  D.~C.,   {Levison} H.~F.,  2009, \mn@doi [\icarus]
  {10.1016/j.icarus.2008.10.025}, \href
  {http://adsabs.harvard.edu/abs/2009Icar..200..292B} {200, 292}

\bibitem[\protect\citeauthoryear{{Birnstiel}, {Dullemond}  \&
  {Brauer}}{{Birnstiel} et~al.}{2010}]{birnstiel10}
{Birnstiel} T.,  {Dullemond} C.~P.,   {Brauer} F.,  2010, \mn@doi [\aap]
  {10.1051/0004-6361/200913731}, \href
  {http://adsabs.harvard.edu/abs/2010A%26A...513A..79B} {513, A79}

\bibitem[\protect\citeauthoryear{{Birnstiel}, {Klahr}  \&
  {Ercolano}}{{Birnstiel} et~al.}{2012}]{birnstiel12}
{Birnstiel} T.,  {Klahr} H.,   {Ercolano} B.,  2012, \mn@doi [\aap]
  {10.1051/0004-6361/201118136}, \href
  {http://adsabs.harvard.edu/abs/2012A%26A...539A.148B} {539, A148}

\bibitem[\protect\citeauthoryear{{Blum} \& {Wurm}}{{Blum} \&
  {Wurm}}{2008}]{blum08}
{Blum} J.,  {Wurm} G.,  2008, \mn@doi [\araa]
  {10.1146/annurev.astro.46.060407.145152}, \href
  {http://adsabs.harvard.edu/abs/2008ARA%26A..46...21B} {46, 21}

\bibitem[\protect\citeauthoryear{{Bukhari Syed}, {Blum}, {Wahlberg Jansson}  \&
  {Johansen}}{{Bukhari Syed} et~al.}{2017}]{bukhari17}
{Bukhari Syed} M.,  {Blum} J.,  {Wahlberg Jansson} K.,   {Johansen} A.,  2017,
  \mn@doi [\apj] {10.3847/1538-4357/834/2/145}, \href
  {http://adsabs.harvard.edu/abs/2017ApJ...834..145B} {834, 145}

\bibitem[\protect\citeauthoryear{{Ciarletti}, {Levasseur-Regourd}, {Lasue},
  {Statz}, {Plettemeier}, {H{\'e}rique}, {Rogez}  \& {Kofman}}{{Ciarletti}
  et~al.}{2015}]{ciarletti15}
{Ciarletti} V.,  {Levasseur-Regourd} A.~C.,  {Lasue} J.,  {Statz} C.,
  {Plettemeier} D.,  {H{\'e}rique} A.,  {Rogez} Y.,   {Kofman} W.,  2015,
  \mn@doi [\aap] {10.1051/0004-6361/201526337}, \href
  {http://adsabs.harvard.edu/abs/2015A%26A...583A..40C} {583, A40}

\bibitem[\protect\citeauthoryear{{Eisenhour}}{{Eisenhour}}{1996}]{eisenhour96}
{Eisenhour} D.~D.,  1996, \mn@doi [Meteoritics and Planetary Science]
  {10.1111/j.1945-5100.1996.tb02019.x}, \href
  {http://adsabs.harvard.edu/abs/1996M\%26PS...31..243E} {31, 243}

\bibitem[\protect\citeauthoryear{{G{\"u}ttler}, {Blum}, {Zsom}, {Ormel}  \&
  {Dullemond}}{{G{\"u}ttler} et~al.}{2010}]{guttler10}
{G{\"u}ttler} C.,  {Blum} J.,  {Zsom} A.,  {Ormel} C.~W.,   {Dullemond} C.~P.,
  2010, \mn@doi [\aap] {10.1051/0004-6361/200912852}, \href
  {http://adsabs.harvard.edu/abs/2010A%26A...513A..56G} {513, A56}

\bibitem[\protect\citeauthoryear{{Hayashi}}{{Hayashi}}{1981}]{hayashi81}
{Hayashi} C.,  1981, \mn@doi [Progress of Theoretical Physics Supplement]
  {10.1143/PTPS.70.35}, \href
  {http://adsabs.harvard.edu/abs/1981PThPS..70...35H} {70, 35}

\bibitem[\protect\citeauthoryear{Hitti \& Bernacki}{Hitti \&
  Bernacki}{2013}]{hitti13}
Hitti K.,  Bernacki M.,  2013, \mn@doi [Applied Mathematical Modelling]
  {http://dx.doi.org/10.1016/j.apm.2012.11.018}, 37, 5715

\bibitem[\protect\citeauthoryear{{Johansen}, {Youdin}  \& {Mac Low}}{{Johansen}
  et~al.}{2009}]{johansen09}
{Johansen} A.,  {Youdin} A.,   {Mac Low} M.-M.,  2009, \mn@doi [\apjl]
  {10.1088/0004-637X/704/2/L75}, \href
  {http://adsabs.harvard.edu/abs/2009ApJ...704L..75J} {704, L75}

\bibitem[\protect\citeauthoryear{{Johansen}, {Blum}, {Tanaka}, {Ormel},
  {Bizzarro}  \& {Rickman}}{{Johansen} et~al.}{2014}]{johansen14}
{Johansen} A.,  {Blum} J.,  {Tanaka} H.,  {Ormel} C.,  {Bizzarro} M.,
  {Rickman} H.,  2014, \mn@doi [Protostars and Planets VI]
  {10.2458/azu_uapress_9780816531240-ch024}, \href
  {http://adsabs.harvard.edu/abs/2014prpl.conf..547J} {pp 547--570}

\bibitem[\protect\citeauthoryear{{Johansen}, {Mac Low}, {Lacerda}  \&
  {Bizzarro}}{{Johansen} et~al.}{2015}]{johansen15}
{Johansen} A.,  {Mac Low} M.-M.,  {Lacerda} P.,   {Bizzarro} M.,  2015, \mn@doi
  [Science Advances] {10.1126/sciadv.1500109}, \href
  {http://adsabs.harvard.edu/abs/2015SciA....115109J} {1, 15109}

\bibitem[\protect\citeauthoryear{{Kofman} et~al.,}{{Kofman}
  et~al.}{2015}]{kofman15}
{Kofman} W.,  et~al., 2015, \mn@doi [Science] {10.1126/science.aab0639}, \href
  {http://adsabs.harvard.edu/abs/2015Sci...349b0639K} {349, 020639}

\bibitem[\protect\citeauthoryear{{Lorek}, {Gundlach}, {Lacerda}  \&
  {Blum}}{{Lorek} et~al.}{2016}]{lorek16}
{Lorek} S.,  {Gundlach} B.,  {Lacerda} P.,   {Blum} J.,  2016, \mn@doi [\aap]
  {10.1051/0004-6361/201526565}, \href
  {http://adsabs.harvard.edu/abs/2016A%26A...587A.128L} {587, A128}

\bibitem[\protect\citeauthoryear{{Massironi} et~al.,}{{Massironi}
  et~al.}{2015}]{massironi15}
{Massironi} M.,  et~al., 2015, \mn@doi [\nat] {10.1038/nature15511}, \href
  {http://adsabs.harvard.edu/abs/2015Natur.526..402M} {526, 402}

\bibitem[\protect\citeauthoryear{{Mottola} et~al.,}{{Mottola}
  et~al.}{2015}]{mottola15}
{Mottola} S.,  et~al., 2015, \mn@doi [Science] {10.1126/science.aab0232}, \href
  {http://adsabs.harvard.edu/abs/2015Sci...349b0232M} {349, 020232}

\bibitem[\protect\citeauthoryear{{Nesvorn{\'y}}, {Youdin}  \&
  {Richardson}}{{Nesvorn{\'y}} et~al.}{2010}]{nesvorny10}
{Nesvorn{\'y}} D.,  {Youdin} A.~N.,   {Richardson} D.~C.,  2010, \mn@doi [\aj]
  {10.1088/0004-6256/140/3/785}, \href
  {http://adsabs.harvard.edu/abs/2010AJ....140..785N} {140, 785}

\bibitem[\protect\citeauthoryear{{Noll}, {Grundy}, {Stephens}, {Levison}  \&
  {Kern}}{{Noll} et~al.}{2008}]{noll08}
{Noll} K.~S.,  {Grundy} W.~M.,  {Stephens} D.~C.,  {Levison} H.~F.,   {Kern}
  S.~D.,  2008, \mn@doi [\icarus] {10.1016/j.icarus.2007.10.022}, \href
  {http://adsabs.harvard.edu/abs/2008Icar..194..758N} {194, 758}

\bibitem[\protect\citeauthoryear{{P{\"a}tzold} et~al.,}{{P{\"a}tzold}
  et~al.}{2016}]{patzold16}
{P{\"a}tzold} M.,  et~al., 2016, \mn@doi [\nat] {10.1038/nature16535}, \href
  {http://adsabs.harvard.edu/abs/2016Natur.530...63P} {530, 63}

\bibitem[\protect\citeauthoryear{{Safronov}}{{Safronov}}{1969}]{safronov69}
{Safronov} V.~S.,  1969, {Evoliutsiia Doplanetnogo Oblaka (English transl.:
  Evolution of the Protoplanetary Cloud and Formation of Earth and the Planets,
  NASA Tech. Transl. F-677, Jerusalem: Israel Sci. Transl., 1972)}

\bibitem[\protect\citeauthoryear{{Sierks} et~al.,}{{Sierks}
  et~al.}{2015}]{sierks15}
{Sierks} H.,  et~al., 2015, \mn@doi [Science] {10.1126/science.aaa1044}, \href
  {http://adsabs.harvard.edu/abs/2015Sci...347a1044S} {347, aaa1044}

\bibitem[\protect\citeauthoryear{{Simon}, {Armitage}, {Li}  \&
  {Youdin}}{{Simon} et~al.}{2016}]{simon16}
{Simon} J.~B.,  {Armitage} P.~J.,  {Li} R.,   {Youdin} A.~N.,  2016, \mn@doi
  [\apj] {10.3847/0004-637X/822/1/55}, \href
  {http://adsabs.harvard.edu/abs/2016ApJ...822...55S} {822, 55}

\bibitem[\protect\citeauthoryear{{Wahlberg Jansson} \& {Johansen}}{{Wahlberg
  Jansson} \& {Johansen}}{2014}]{wahlberg14}
{Wahlberg Jansson} K.,  {Johansen} A.,  2014, \mn@doi [\aap]
  {10.1051/0004-6361/201424369}, \href
  {http://adsabs.harvard.edu/abs/2014A%26A...570A..47W} {570, A47}

\bibitem[\protect\citeauthoryear{{Wahlberg Jansson}, {Johansen}, {Bukhari Syed}
   \& {Blum}}{{Wahlberg Jansson} et~al.}{2017}]{wahlberg17}
{Wahlberg Jansson} K.,  {Johansen} A.,  {Bukhari Syed} M.,   {Blum} J.,  2017,
  \mn@doi [\apj] {10.3847/1538-4357/835/1/109}, \href
  {http://adsabs.harvard.edu/abs/2017ApJ...835..109W} {835, 109}

\bibitem[\protect\citeauthoryear{{Weidenschilling}}{{Weidenschilling}}{1977}]{%
weidenschilling77}
{Weidenschilling} S.~J.,  1977, \mn@doi [\mnras] {10.1093/mnras/180.1.57},
  \href {http://adsabs.harvard.edu/abs/1977MNRAS.180...57W} {180, 57}

\bibitem[\protect\citeauthoryear{{Whipple}}{{Whipple}}{1972}]{whipple72}
{Whipple} F.~L.,  1972, in {Elvius} A.,  ed., From Plasma to Planet. p.~211

\bibitem[\protect\citeauthoryear{{Youdin} \& {Goodman}}{{Youdin} \&
  {Goodman}}{2005}]{youdin05}
{Youdin} A.~N.,  {Goodman} J.,  2005, \mn@doi [\apj] {10.1086/426895}, \href
  {http://adsabs.harvard.edu/abs/2005ApJ...620..459Y} {620, 459}

\bibitem[\protect\citeauthoryear{{Zsom} \& {Dullemond}}{{Zsom} \&
  {Dullemond}}{2008}]{zsom08}
{Zsom} A.,  {Dullemond} C.~P.,  2008, \mn@doi [\aap]
  {10.1051/0004-6361:200809921}, \href
  {http://adsabs.harvard.edu/abs/2008A%26A...489..931Z} {489, 931}

\makeatother
\end{thebibliography}

\appendix

\section{Enclosed mass and potential energy of sub-cloud $\lowercase{i}$ in the radially resolved pebble cloud model}\label{app:A}

In the radially resolved model described in the main text the enclosed mass for each sub-cloud $i$, $M_{i,\t{enc}}$, can be written as

\begin{align}
 M_{i,\t{enc}} &= \sum_{k=1}^{N_p}m_{ik,\t{enc}} + M_\t{c}\ , \label{eq:mEnc}\\
 m_{ik,\t{enc}} &= \left\{\begin{array}{lr}
  \left(\frac{R_i}{R_k}\right)^3m_k\ , & R_i\leq R_k \\
  m_k\ , & R_i>R_k
 \end{array} \right. \label{eq:mEncik}
\end{align}

\noindent where $m_i$ and $R_i$ is the mass and radius of sub-cloud $i$ and $M_\t{c}$ is the mass of any solid core. We get the total potential energy for sub-cloud $i$, $U_i$, from Poisson's equation,

\begin{align}
 U_i &= \sum_{k=1}^{N_p}U_{ik} - \frac{3GM_\t{c}m_i}{2R_i} \ , & \label{eq:appU} \\
 U_{ik} &= \int_0^{R_i}\Phi_{ik}(r)\d m_i(r)\ , \quad \d m_i(r) = 4\pi\rho_ir^2\d r \\
 \Phi_{ik}(r) &= \left\{\begin{array}{lr}
     \frac{2}{3}\pi G\rho_k\left(r^2-3R_k^2\right)\ , & r\leq R_k \\
     -\frac{Gm_k}{r}\ , & r>R_k \\
     -\frac{4}{3}\pi G\rho_ir^2\ , & i=k
    \end{array} \right.   \\
 \Longrightarrow U_{ik} &= \left\{\begin{array}{lc}
     -\frac{3Gm_im_k}{2R_k}\left(1-\frac{1}{5}\left(\frac{R_i}{R_k}\right)^2\right)\ , & R_i\leq R_k \\
     -\frac{3Gm_im_k}{2R_i}\left(1-\frac{1}{5}\left(\frac{R_k}{R_i}\right)^2\right)\ , & R_i>R_k \\
     -\frac{6Gm_i^2}{5R_i}\ , & i = k
    \end{array} \right. \label{eq:appUik}
\end{align}

\medskip
\noindent where $\rho_i$ is the mass density inside sub-cloud $i$. Compared to the homogeneous model in \citetalias{wahlberg14} and \citetalias{wahlberg17} the equilibrium radii in the extended model cannot be found analytically, except for the smallest and largest sub-cloud. As described in the main text \srefp{sec:energies} this is solved numerically by keeping track of the potential energies of all sub-clouds in a sorted array. For a cloud in virial equilibrium we have

\begin{align}
  2E_{i,\t{tot}} &= U_{i,\t{eq}}\ , \label{eq:appVir}
\end{align}

\noindent where $E_{i,\t{tot}}$ is the current total energy of sub-cloud $i$. If $i$ has an equilibrium radius within all other sub-clouds ($R_{i,\t{eq}}<R_k$, $\forall k$) it can be calculated through \eref{eq:appU} and \eref{eq:appVir},

\begin{align}
 U_{i,\t{eq}} &= -\sum_{k\neq i}\frac{3Gm_im_k}{2R_k}\left(1-\frac{1}{5}\left(\frac{R_{i,\t{eq}}}{R_k}\right)^2\right) -\frac{6Gm_i^2}{5R_{i,\t{eq}}} \nonumber \\
              & \quad - \frac{3GM_\t{c}m_i}{2R_{i,\t{eq}}} \nonumber \\
 &= \underbrace{\frac{3Gm_i}{10}\sum_{k\neq i}\frac{m_k}{R_k^3}}_{\Sigma_1}R_{i,\t{eq}}^2 - \underbrace{3Gm_i\left(\frac{M_\t{c}}{2}+\frac{2m_i}{5}\right)}_{\Sigma_2}\frac{1}{R_{i,\t{eq}}} \nonumber \\
 & \quad -\underbrace{\frac{3Gm_i}{2}\sum_{k\neq i}\frac{m_k}{R_k}}_{\Sigma_3}\ , \\
 \Longrightarrow 0 &= R_{i,\t{eq}}^3 - \frac{2E_{i,\t{tot}}+\Sigma_3}{\Sigma_1}R_{i,\t{eq}} - \frac{\Sigma_2}{\Sigma_1}\ , \label{eq:Req_in}
\end{align}

\noindent and solve for $R_{i,\t{eq}}$. Similarly we can find $R_{i,\t{eq}}$ if $i$ is larger than the radius of all other sub-clouds, ($R_{i,\t{eq}}>R_k$, $\forall k$),

\begin{align}
  U_{i,\t{eq}} &= -\sum_{k\neq i}\frac{3Gm_im_k}{2R_{i,\t{eq}}}\left(1-\frac{1}{5}\left(\frac{R_k}{R_{i,\t{eq}}}\right)^2\right) - \frac{6Gm_i^2}{5R_{i,\t{eq}}} \nonumber \\
               & \quad - \frac{3GM_\t{c}m_i}{2R_{i,\t{eq}}} \nonumber \\
  &= -\underbrace{3Gm_i\left(\sum_{k\neq i}\frac{m_k}{2}+\frac{2m_i}{5}+\frac{M_\t{c}}{2}\right)}_{\Sigma_4}\frac{1}{R_{i,\t{eq}}} \nonumber \\
               & \quad + \underbrace{\frac{3Gm_i}{10}\sum_{k\neq i}m_kR_k^2}_{\Sigma_5}\cdot\frac{1}{R_{i,\t{eq}}^3}\ , \\
\Longrightarrow 0 &= R_{i,\t{eq}}^3 + \frac{\Sigma_4}{2E_{i,\t{tot}}}R_{i,\t{eq}}^2 - \frac{\Sigma_5}{2E_{i,\t{tot}}}\ , \label{eq:Req_out}
\end{align}

\noindent and solve for $R_\t{i,eq}$.

\bsp	
\label{lastpage}
\end{document}